\documentclass[conference]{IEEEtran}
\usepackage{amsmath,amssymb,amsfonts}

\usepackage{cite}
\usepackage{graphicx}
\usepackage{psfrag}
\usepackage{subfigure}
\usepackage{amsmath}
\usepackage{array}
\usepackage{algorithmicx}
\usepackage{algpseudocode}
\usepackage{setspace}
\usepackage{blindtext}
\usepackage{url}
\usepackage{epstopdf}
\usepackage{verbatim}
\usepackage{color}
\usepackage{amsmath}
\usepackage{bm}
\usepackage{multirow}
\usepackage[table,xcdraw]{xcolor}
\usepackage{booktabs}
\usepackage{makecell}
\usepackage{ntheorem}
\usepackage{textcomp}
\usepackage{amsmath}

\newtheorem{Lem}{Lemma}

\newtheorem{Prob}{Problem}

\newcommand{\RNum}[1]{\uppercase\expandafter{\romannumeral #1\relax}}

\theoremseparator{:}

\title{A Novel Cross-band CSI Prediction Scheme for Multi-band Fingerprint based Localization}
\author{Ruihao Yuan, Kaixuan Huang, Yuru Duan, Shunqing Zhang\\
Shanghai Institute for Advanced Communication and Data Science, \\
Shanghai University, Shanghai, 200444, China\\
Email: \{yuan980728, xuan1999, sehunyara999,  shunqing\}@shu.edu.cn}

\begin{document}
\maketitle

\begin{abstract}

Because of the advantages of computation complexity compared with traditional localization algorithms, fingerprint based localization is getting increasing demand. Expanding the fingerprint database from the frequency domain by channel reconstruction can improve localization accuracy. However, in a mobility environment, the channel reconstruction accuracy is limited by the time-varying parameters. In this paper, we proposed a system to extract the time-varying parameters based on space-alternating generalized expectation maximization
(SAGE) algorithm, then used variational auto-encoder (VAE) to reconstruct the channel state information on another channel. The proposed scheme is tested on the data generated by the deep-MIMO channel model. Mathematical analysis for the viability of our system is also shown in this paper. 
\end{abstract}


\IEEEpeerreviewmaketitle
\section{Introduction} \label{sect:intro}

Integrated sensing and communication (ISAC) has been identified as a key enabling technology for the sixth generation (6G) mobile communication networks \cite{6GISAC}. In addition to the conventional communication capability, wireless signals have been applied to sense the physical environment via signal fingerprints. For example, the cellular signal fingerprints have been adopted to track the trajectories of vehicles in the outdoor environment \cite{10157900}, and the WiFi signal fingerprints have been used to sense the motion behavior of robots in the indoor environment\cite{8275781}.

A straightforward and efficient idea to improve the sensing capability is to merge different signal fingerprints from different frequency bands together \cite{7990593}. As reported in \cite{8423070}, indoor localization accuracy can be improved from 1.75 meters to 0.95 meters by leveraging multi-band signal fingerprints. A similar concept has been extended to millimeter-wave sensing as well \cite{souzandeh2023classification}, where the sensing range can be improved from 16 cm to 3.25 m, if the radiation characteristics from multiple signal bands are jointly utilized. By utilizing this type of {\em fingerprint diversity} in different frequency bands, the sensing performance improvement is promising at the expense of sampling multiple frequency bands. In order to reduce the sampling complexity of multi-band signal fingerprints, we propose to learn the cross-band nonlinear correlation offline and infer the cross-band channel state information (CSI) online using variational auto-encoder (VAE) network in \cite{yuan2023variational}, and eventually improves the localization accuracy to 0.918 meters in the indoor environment. 

The above multi-band signal fingerprint based sensing or localization schemes, however, rely on the assumption that all the sensing or localization entities share the same mobility environment. Once the mobility conditions for different entities are {\em NOT} the same, we can not share the same projection method on the same latent space through machine learning. Another possibility is to take the mobility condition as an input and recover the CSI condition accordingly. For example, an extended Kalman filter (EKF) based channel estimation algorithm has been proposed in \cite{8424581} to track the temporal correlations using an iterative detector decoder, and in \cite{shi2021unified}, the super-resolution based deep learning scheme has been proposed to characterize the non-linear temporal correlations. These types of methods require reference signals in different frequency bands to obtain fingerprints, while the implementation complexity is still significant. In order to keep the fingerprint diversity with affordable complexity, we propose a cross-band CSI prediction scheme with different mobility conditions and present a VAE enabled multi-band fingerprint based localization algorithm to answer the following three questions.

\begin{itemize}
    \item {\em How to efficiently characterize the mobility conditions?} 
    Instead of directly estimating the velocity of each entity as explained in \cite{10128162}, we propose to calculate fading coefficients of multiple propagation paths using weighted multiple signal classification (MUSIC) algorithm \cite{10000625} and extract the mobility conditions via a modified space-alternating generalized expectation (SAGE) \cite{article1} scheme. Through this approach, we are able to efficiently track small-scale mobility variations by observing small-scale multi-path fading coefficients.      
    \item {\em How to figure out the cross-band mobility independent features?} 
    In addition to the previous VAE network \cite{yuan2023variational} for cross-band CSI inference, we propose to apply a neural network-enabled mobility removal block before the VAE architecture. By using this type of structure, we can obtain the cross-band mobility independent features and reconstruct multi-band signal fingerprints accordingly.    
    \item {\em How to utilize fingerprint diversity for more accurate localization performance?} By jointly utilizing the mobility extraction scheme and modified VAE architecture, we can incorporate the fingerprint diversity via cross-band CSI reconstruction. Through this type of data augmentation approach, we are able to achieve localization accuracy up to 1.4535 meters in the mobility case, which performs much better than the single band baselines and is close to the multi-band performance bound.
\end{itemize}

The rest of this paper is organized as follows. In Section~\ref{sect:sys}, we introduce some preliminary information and the channel model we used in this paper. The problem formulation is discussed in Section~\ref{sect:prob}. The detailed algorithm procedure and the whole scheme are provided in Section~\ref{sect:nndesign}. In Section~\ref{sect:experiment}, we present our channel reconstruction experimental results, and the result of an application of localization with the estimated channel information is shown. The concluding remarks are given in  Section~\ref{sect:conc}.

\section{System Model and Preliminary} \label{sect:sys}


\subsection{System Model}
\label{subsect:prob}
Consider a multi-band orthogonal frequency division multiplexing
(OFDM) enabled transmission system, a transmitter with a single antenna and a receiver with multiple antennas are set up as the tracking system. There is a Line-of-sight (LoS) path existing between the transmitter and receiver.
The wireless channel, due to the multi-path effect, has the following measurement at channel $n$, time $t$, sub-carrier $f$, and antenna $a$, which can be expressed as:
\begin{eqnarray}
\label{eq:model}
\mathbf{H}_n(t,f,a,\mathcal{L}_t)=\sum_{l = 1}^{L}P_{l}^n(t,f,a,\mathcal{L}_t)+N(t,f,a,\mathcal{L}_t) \nonumber
\\= \sum_{l = 1}^{L}\alpha_{l}^n(t,f,a,\mathcal{L}_t)e^{-j2\pi f\tau_{l}(t,f,a,\mathcal{L}_t)} +N(t,f,a,\mathcal{L}_t)
\end{eqnarray}

where $L$ is the total number of multi-path components (MPCs), $P_{l}$ is the signal of the $l$-th path, $\alpha_{l}$ and $\tau$ are the complex attenuation factor and propagation delay of the $l$-th path, respectively. $N(t,f,a,\mathcal{L}_t)$ is the zero-mean unit complex Gaussian white noise. 

Taking $\mathbf{H}_n(0,0,0,0)$ as a reference, the phase of $\mathbf{H}_n(t,f,a,\mathcal{L}_t)$ can be expressed as\cite{10.1145/3210240.3210314}: 
\begin{eqnarray}
\label{eq:DS model}
f{\tau_{l}} (t,f,\alpha,\mathcal{L}_t) \approx f_c \tau_{l} + \Delta f_i \tau_{l} + f_c \Delta s_k \phi_{l} - f_{D_{l}}\Delta t_i 
\end{eqnarray}
In this, $f_c$ is the central carrier frequency, $\Delta t_i$, $\Delta f_i$, $\Delta s_k$ are the differences of time, frequency and spatial position between $\mathbf{H}(t,f,a,\mathcal{L}_t)$ and $\mathbf{H}(0,0,0,0)$, respectively. $\tau_{l}$, $\phi_{l}$ and $f_{D_{l}}$ are the ToF, AoA and DFS of the $l$-th path, respectively. 
Let $t = 0,1,...,T - 1; f = 0,1,...,F-1; a = 0,1,...,S-1$, where $T$, $F$, $S$ represent for the number of packets, sub-carriers, and antennas respectively. The location of UE while sampling is denoted by $\mathcal{L}_t$. We denote $m = (t,f,a,\mathcal{L}_t)$ for brevity. Note that the term $f_c \tau_{l}$ in the formulation is the same in all measurements and can be merged into the complex attenuation. 

Through the operation described above, record the signal parameters of the path as $\theta_{l}(\alpha_{l},\tau_{l},\phi_{l},f_{D_{l}})$, and the first step of tracking is to estimate the multidimensional parameters $\theta$ of the reflected signal of the target. After removing the DS from $\theta_{l}$, the parameter becomes a static parameter. We denoted the static parameter as $\hat{\theta}_{l} = (\alpha_{l},\tau_{l},\phi_{l})$, Every multipath component has its own $\theta_{l}$, we could denote the $P_{l}(m)$ under specific $\theta_{l}$ as $P_{l}(m;\theta_{l})$. Similarly, the static path is denoted as $\hat{P}_{l}(m;\hat{\theta}_{l})$ The expression of $\mathbf{H}(m)$ contains the DS, which is a dynamic parameter under different speed while the target is moving. After removing the DS, the channel would be recovered to a static channel. Thus, we need to generate the parameters on every path and extract the static and dynamic parameters from them.

\begin{figure}
\centering
\includegraphics[width = 3 in,height=2.3in]{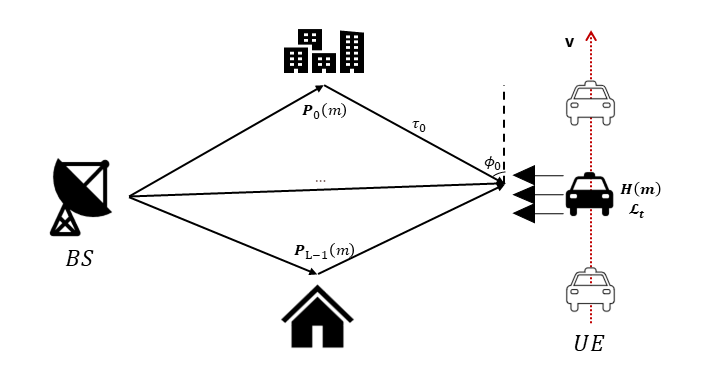}
\caption{The system model.}
\label{fig:sys}
\end{figure}

The whole system is shown in Fig.\ref{fig:sys}. There is a LoS path and several NLoS paths between BS and UE. UE is moving at a speed $v$ which is unknown, and the OFDM signals are sampled from UE. Based on different positions, By collecting channel fading coefficients from different frequency bands together, we construct the localization database according to the following format.
\begin{eqnarray}
\mathcal{DB} = \left\{\left(\mathcal{L}, {\overline{\mathbf{H}}_1}(t,f,a,\mathcal{L}), \ldots, \overline{{\mathbf{H}}}_{N_B}(t,f,a,\mathcal{L})\right)\right\},
\end{eqnarray}
$ \forall n \in [1, \ldots, N_B]$, denotes the measured channel responses of the $n$-th frequency band. After sampling the multi-band CSI, the data could be matched with  $\mathcal{DB}$ to generate the location. $\overline{\mathbf{H}}$ is the static channel.
 
\section{Problem Formulation}
\label{sect:prob}

The localization problem is formulated by the following:

\begin{Prob}[\em MSE Minimization] The MSE minimization problem on channel $n'$ is formulated by,
\begin{eqnarray}
\underset{\mathcal{I}_n(\cdot)}{\textrm{minimize}} && \frac{1}{T}\sum_{t=1}^{T} \|\mathcal{L}_t-\hat{\mathcal{L}}_{t}\|_2
\label{eqn:mini1}\\
\textrm{subject to}
&&\hat{\mathcal{L}}_{t} = \Gamma(\mathcal{DB},[\overline{\mathbf{H}}_{n}(m),\mathbf{H }_{n'}^{\star}(m)])\\
&&\mathbf{H}_{n'}^{\star}(m) = argmin \nonumber\\
&&\frac{1}{M}\sum_{m=1}^{M} \|\overline{\mathbf{H}}_{n'}(m)-\tilde{\mathbf{H}}_{n'}(m)\|_2\\
 && \overline{\mathbf{H}}_{n'}(m) = \sum_{l=1}^{L}\overline{P}^{n'}_{l}(m;\theta_{l}),\\
 && \tilde{\mathbf{H}}_{n'}(m) = \sum_{l=1}^{L}\tilde{P}^{n'}_{l}(m;\theta_{l}),\\
&& n, n' \in N_{sc}
\end{eqnarray}
where $\|\cdot\|_2$ denotes the $l_2$ norm of the inner vector. $\Gamma(\cdot)$ is the localization function. $\tilde{\mathbf{H}}_{n'}(m)$ is the predicted static CSI on channel $n'$. $\mathbf{H }_{n'}^{\star}(m)$ is the most optimized static channel on channel $n'$.
\end{Prob}

The minimization problem is hard to solve in a mobility environment, thus we turn this problem into the following two problems.
\begin{Prob}[\em Mobility Parameters Estimation]
\label{Prob:2}
\begin{eqnarray}
\underset{\mathcal{J}_n(\cdot)}
{\textrm{minimize}} && \sum_{m=1}^{M}\|\mathbf{H}_{n'}(m)-\hat{\mathbf{H}}_{n'}(m)\|_2 
\label{eqn:2}\\
\textrm{subject to} && \hat{\mathbf{H}}_{n'}(m) = \mathcal{J}_n(\mathbf{H}_{n'}(m)),\\
&& \hat{\mathbf{H}}_{n'}(m) = \sum_{l=1}^{L}\hat{P}^{n'}_{l}(m;\theta_{l}),\label{equ:11}
\end{eqnarray}
where $\mathcal{J}_n(\cdot)$ is the function to estimate the parameters on mobility channel $\mathbf{H}_n(m)$. 
\end{Prob}

The parameter of the multi-path component should be determined to maximize the MLE problem. After estimating the parameters, we recovered the static channel by removing the dynamic parameters.
\begin{Prob}[\em Channel Reconstruction Error Minimization] 
\label{Prob3}
The channel estimation problem is also an MSE minimization problem, as we have introduced in \cite{yuan2023variational}.
\begin{eqnarray}
\underset{\mathcal{G}_n(\cdot),\mathcal{X}_n(\cdot)}{\textrm{minimize}} && \frac{1}{M}\frac{1}{L}\sum_{m=1}^{M}\sum_{l=1}^{L} \|\overline{{P}}^{n'}_{l}(m) -\tilde{{P}}_{l}^{n'}(m)\|_2
\label{eqn:mini2}\\
\textrm{subject to} 
&& \tilde{{P}}^{n'}_{l}(m;\hat{\theta}_{l}) = \mathcal{G}_n\left(\tilde{P}^{n}_{l}(m;\hat{\theta}_{l})\right),\label{eq:begin}\\
&&\tilde{P}^{n}_{l}(m;\theta_{l}) = \mathcal{X}_{n}({\hat{P}^{n}_{l}}(m;\theta_{l})),\\
&&\overline{P}^{n'}_{l}(m;\theta_{l}) = \mathcal{X}_{n'}({{P}^{n'}_{l}}(m;\theta_{l})),\\
&&\hat{P}^{n'}_{l}(m;\theta_{l}) = \hat{\mathcal{X}}^{-1}_{n'}({\tilde{P}^{n'}_{l}}(m;\theta_{l}))\label{eq:end}
\end{eqnarray}
where $\mathcal{G}_n(\cdot)$ is the function that maps static path on channel $n$ to channel $n'$, and $\mathcal{X}_n(\cdot)$ is the function that departs the static parameters and dynamic parameters, and recover the static CSI. $\overline{{P}}^{n'}_{l}(m)$ is the truth value of the signal on path $l$ on channel $n'$, and $\tilde{{P}}^{n'}_{l}(m)$ is the predicted value.
\begin{Lem} \label{lem:equ}
Problem 1 and Problem 3 are linear connected and when the errors of the two problems are minimized, the error of problem 2 is also minimized.
\begin{IEEEproof}
Please refer to Appendix~\ref{appen:elbo} for the proof.
\end{IEEEproof}
\end{Lem}
\end{Prob}

\section{Solution Based on SAGE and VAE}
\begin{figure*}[ht]
    \centering
    \includegraphics[width=7 in]{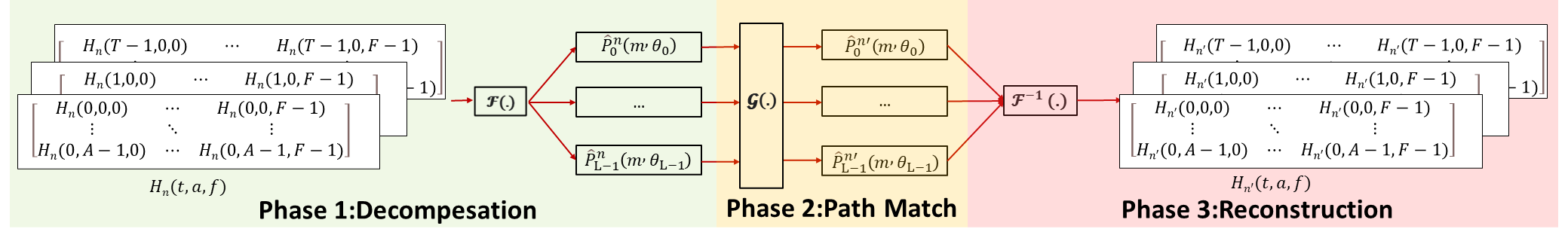}
    \caption{The whole procedure of the system.}
    \label{fig:procedure}
\end{figure*}

  The whole system is shown in Fig.\ref{fig:procedure}. The parameter estimation procedure from  $H(m)$ to static parameters $\hat{P}_{l}(m)$ is denoted by $\mathcal{F}(\cdot)$. The prediction function that reconstructed the static path components on another channel is denoted by $\mathcal{G}(\cdot)$. After the prediction of the static path on the channel $n'$, the time-varying is recovered from the static path and all the paths are summed up to reconstruct the time-varying channel. $\mathcal{F}(\cdot)$ includes 
$\mathcal{J}(\cdot)$ and $\mathcal{X}(\cdot)$ in Chapter \ref{Prob3}, which represent the parameter estimation and removing the mobility parameters.  
\label{sect:nndesign}
\subsection{Modified SAGE Algorithm for Parameters Estimation}

The channel estimation problem is to maximize the optimization problem in \textbf{Problem \ref{Prob:2}}, and this problem is non-linear and has no closed-form solution. So we applied space-alternating generalized expectation maximization
(SAGE) algorithm\cite{dempster1977maximum} to solve this problem. Inspired by \cite{10000625}, before applying the SAGE algorithm, we use the well-known weighted MUSIC algorithm to estimate the coarse evaluation of the parameters and the coarse evaluation is used as the initial value for the SAGE algorithm.

SAGE is widely used in literatures to estimate the direction of departure
(DoD), the direction of arrival (DoA) time of flight and DS of the MPCs\cite{10.1145/3210240.3210314}\cite{Yin2016PerformanceCO}\cite{6901837}. SAGE is an extension of the Expectation Maximization (EM) algorithm. In each iteration of the SAGE algorithm, only a subset of the components are updated, while keeping the estimates of the other components fixed. Thus, we can divide the estimate of $\Theta$ into multiple estimates of individual parameters. The parameters of each path are optimized in turn. Specifically, for the $l$-$th$ path on channel $n$, the expectation step is to construct the signal on $l$-$th$ path:
\begin{eqnarray}
\xi^n_{l}(m;{\Theta}') = \mathbf{H}_n(m)-\sum_{\substack{l'=1, l'\neq l}}^{L}\overline{{P}^{n}_{l}}(m;{{\theta}}'_{l'})
\end{eqnarray}
where  $\Theta'$ is the parameter estimated in the last iteration. Because all the signals on $L$ paths are added up on the observation value $\mathbf{H}_n(m)$, we removed the other paths' signals to restore the $l$-$th$ path. And $\xi^n_{l}(m;{\Theta}')$ is the expectation equation on E-step. 

Then, the maximization step is given in\cite{fleury1999channel}:
\begin{eqnarray}
&&{\hat{\tau}_{l}}'' = argmax_\tau\{|z(\hat{\tau},\hat{\phi}'_{l},\hat{f}'_{D_{l}};\xi^n_{l}(m;{\Theta}')))|\}\\
&&{\hat{\phi}_{l}}'' = argmax_\phi\{|z(\hat{\tau}''_{l},{\phi},\hat{f}'_{D_{l}};\xi^n_{l}(m;{\Theta}'))|\}\\
&&\hat{f}''_{D_{l}} = argmax_{f_D}\{|z(\hat{\tau}''_{l},{\hat{\phi}''_{l}},{f}_D;\xi^n_{l}(m;{\Theta}'))|\}\\
&&\hat{a}''_{l} = \frac{1}{M} z(\hat{\tau}''_{l},{\hat{\phi}''_{l}},{f}''_{D_{l}};\xi^n_{l}(m;{\Theta}'))\\
\end{eqnarray}
where,
\begin{eqnarray}
z(\tau,\phi,f_D;\xi^n_{l}) = \sum_{m}^{} e^{2\pi\Delta f_j\tau_{l}}e^{2\pi f_c\Delta s_k\phi_{l}}e^{-2\pi f_{D_{l}}\Delta t_i}\xi^n_{l}(m) 
\end{eqnarray}
Since the MLE of  $\alpha_{l}$ can be derived in a closed form as a function of $\tau_{l}$, $\Phi_{l}$ and $f_{D_{l}}$, it is calculated at the end of each iteration. The SAGE algorithm can be viewed as a grouped coordinate ascent method. 

\subsection{Deep Learning Based Channel Reconstruction}

The channel reconstruction problem is to minimize the optimization problem in \textbf{Problem \ref{Prob3}}. In \cite{yuan2023variational}, we introduced the channel reconstruction method using the VAE network in detail. Based on the hypothesis for our end-to-end approach, the down-link channel can be obtained from the up-link channel. Therefore, we learn the distribution of the down-link channel given the up-link channel (instead of learning the distribution of the up-link channel, as would be the case in traditional VAE). We use the encoder network to learn a lower-dimensional interpretable representation $Z = R^l$ for our target distribution. The decoder then uses the obtained representation to predict the downlink channel. In mobility environments, because of the existence of the time-varying 



Before applying the VAE network we have introduced, we used an FNN network to map the relation between mobility parameters channel $n$ and channel $n'$. the input of the network is the time-varying parameters extracted by the MUSIC-SAGE algorithm, and the output data is the channel mobility parameters on the channel $n'$. The network structure of VAE is the same as our previous work \cite{yuan2023variational}, and the network structure of FNN is listed in Table. 
\begin{table} [ht]
\centering
\caption{Parameters Set on FNN Network.}
\label{tab:CNN}
\footnotesize
\renewcommand\arraystretch{1.5}
	\begin{tabular}{c|c|c|c}  
		\hline  
	Parameter& Value& Parameter &Value  \\ 
        \hline
        input/output layer size& $N_{sc}$ & learning rate & 0.01\\
        \hline
        hidden layer size& FC 128 & active function &ReLU\\
        \hline
        training data size& $N_{sc}\cdot A\cdot T\cdot L$ & optimizer &Adam\\
        \hline
	\end{tabular}
\end{table}

\section{Experiment Results} \label{sect:experiment}
In this section, we tested the performance of our proposed scheme in a mobility environment. In this paper, we firstly compared MUSIC-SAGE with the other two optimization algorithms. {\em Baseline 1: Particle Swarm Optimization(PSO) algorithm\cite{934374}}, PSO algorithm seeks the optimal solution
by simulating the social behaviors of a bird flock. It is widely used on channel estimation\cite{9740477}. {\em Baseline 2: Covariance matrix Adaptation Evolution Strategy (CMA-ES)\cite{hansen2016cma}}, The CMA-ES is a stochastic, or randomized, method of optimization for non-linear, non-convex functions. The two baselines have the same feature as SAGE they can all reduce the search space to accomplish the path match step. We tested the results of the three algorithms with the initial evaluation. We also tested the performance of the MUSIC algorithm alone. Then we tested the channel reconstruction accuracy of our scheme under different velocities. Finally, fingerprint based localization experiment was deployed to estimate the performance of our whole system.

The data of the numerical experiment was generated by the deep-MIMO model\cite{alkhateeb2019deepmimo}. We generated the data from 10 users, and the users  10 users were spatially spaced out, the movement traces of the users were lines. We chose the 60GHz frequency channel to test our algorithm, and we generated 128 subcarrier data, which were divided into two channel bands, each having 64 subcarriers. The number of receive antennas was 3, and the number of transmit antenna was set as 1. The other parameters were set to default. The scenario of the deep-MIMO dataset was 'O1$\_$60'. 

\subsection{Parameters Estimation}
In this case, we tested the mobility parameters estimation accuracy of the three chosen optimization algorithms. As we have mentioned above, the data is generated in an environment of 60 km/h, and we generated 3584 pieces of data packets. In this paper, the criteria of experiment results is the channel coefficient normalized error(CCNE) as shown in Eq. (\ref{NE}).
\begin{eqnarray}
\label{NE}
CCNE = -10\lg{\left(\frac{\|\tilde{\mathbf{X}}-{\mathbf{X}}\|^2}{\|{\mathbf{X}}\|^2}\right)}. 
\end{eqnarray}
$\tilde{\mathbf{X}}$ is the estimated parameter, and ${\mathbf{X}}$ is the truth value of the parameter. The result is shown in Table \ref{tab:parameter}. 

As shown in Table \ref{tab:parameter}, the accuracy of the parameters' reconstruction of MUSIC-SAGE is better than other optimization algorithms. The CCNEs are shown in Table \ref{tab:parameter}. It can be seen that the MUSIC-SAGE algorithm achieves the best performance on parameter estimation, and MUSIC-CMA-ES is better than MUSIC-PSO.
\begin{table} [ht]
\centering
\caption{An Overview Of Mean CCNE for Each Algorithm.}
\label{tab:parameter}
\footnotesize
\renewcommand\arraystretch{1.5}
	\begin{tabular}{cc|cc}  
		\hline  
	 Algorithm&CCNE& Algorithm&CCNE  \\ 
        \hline
        MUSIC-SAGE& 8.8200&MUSIC-PSO& 6.7355\\
        \hline
        MUSIC-CMA-ES& 7.3580& MUSIC& 4.2609\\
        \hline
	\end{tabular}
\end{table}
\subsection{Channel Reconstruction}
\label{channel reconstruction}
\begin{figure}
\centering
\includegraphics[width = 3 in,height=2.3in]{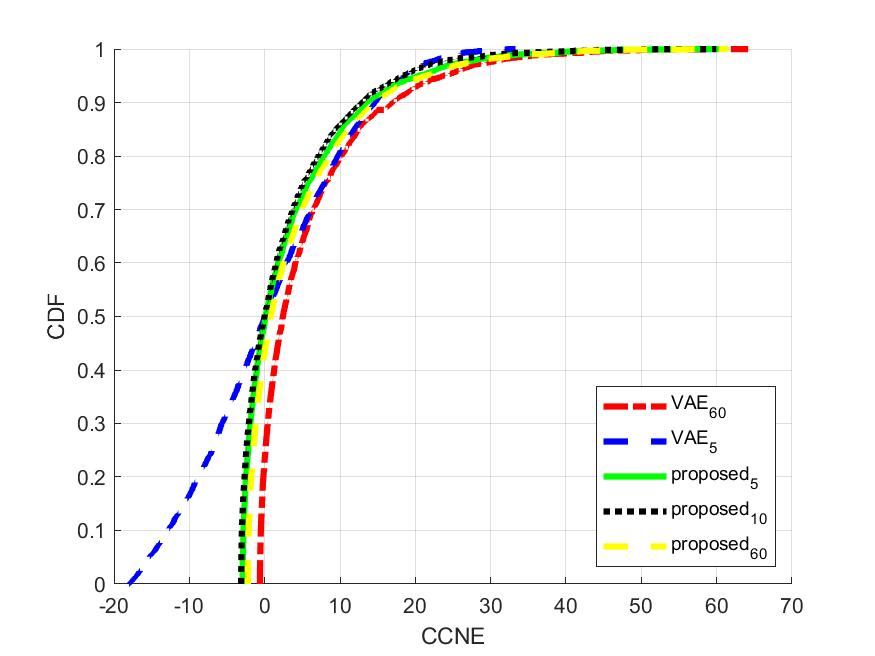}
\caption{CDF of CCNE in channel reconstruction.}
\label{fig:channel reconstruction}
\end{figure}

After the parameters' estimation, we reconstructed every path of the static channel information on channel $n$, then we used the trained VAE model to map each path to channel $n'$. 
After removing the dynamic parameters, we use the VAE network to reconstruct the channel $n'$ directly. We used 28672 pieces of data to train the network, the quantity of the training dataset is 8 times the quantity of the testing dataset, and the training dataset was also generated from deep-MIMO. The results are shown in Fig. \ref{fig:channel reconstruction}. The training data of the VAE network was generated under 60 km/h, while the testing data was generated under 60$\cdot$km/h, 5$\cdot$km/h and the average velocity $\overline{v}$ = 60$\cdot$km/h (the velocity was distributed evenly on 50$\cdot$km/h to 70$\cdot$km/h) The average CCNE was 0.7023, 5.7386 and 3.824, respectively. It can be seen that if the testing data was generated under the same as the training data, the prediction results were quite reliable, however, if the speed was different, there was a huge degradation of the results. 

Then, we applied the SAGE algorithm to our system. It can be seen from the picture that although the testing speeds are far different, the experiment results are quite close. The experiment results show that our proposed system could be applied under any vehicle's speed. The average CCNE of our proposed scheme under the three speeds are 3.3247, 2.8135, and 4.0609, respectively.

\subsection{Localization Results}
\begin{figure}
\centering
\includegraphics[width = 3 in,height=2.3in]{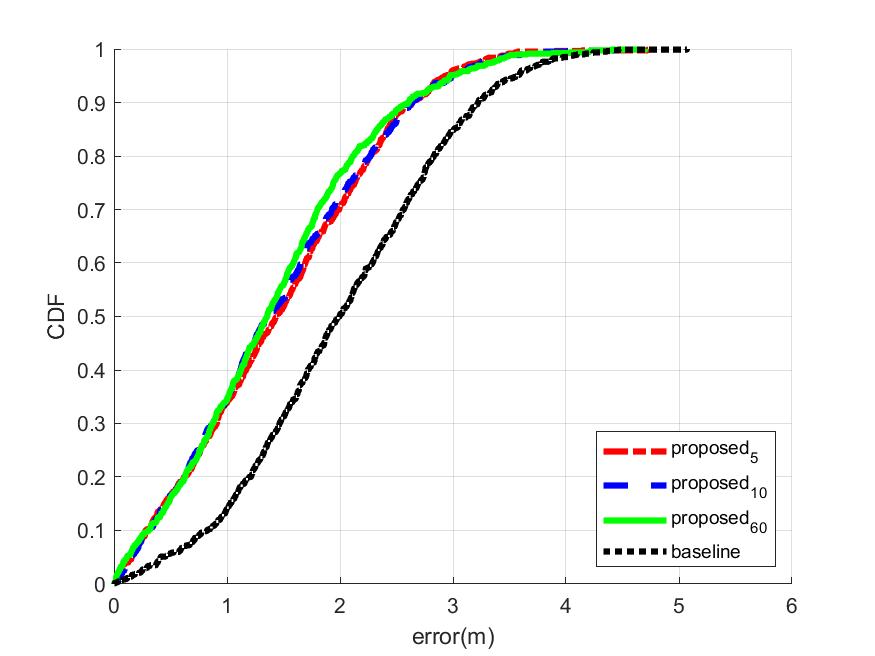}
\caption{CDF of localization error.}
\label{fig:localization}
\end{figure}

We tested the localization accuracy based on the data we reconstructed and sampled. We used the DNN network model to generate the fingerprint database, the parameters of the network and the way splicing channel information are the same as our previous work \cite{yuan2023variational}, and we did a similar experiment which compared the localization accuracy of the original data and the spliced data. Data was spliced by the original data from the channel $n$ and the predicted channel $n'$. The data was also generated from the deep-MIMO database. We chose 16 $\mathcal{L}_t$ as reference points (RP) and sampled 10000 pieces of CSI from each to build the training dataset, and the number of testing points (TP) was 8. We sampled 100 pieces of CSI from TP to validate our work. As shown in Fig. \ref{fig:localization}, the baseline results are multi-band performance bound and single-band performance bound. It can be seen that by expanding the localization resource from the frequency domain through our proposed scheme, the localization accuracy is between using the upper bound and lower bound. The mean error of the four localization results is 1.1094m, 1 4535m, 1.5325m, and 1.8356m, respectively. By expanding the information on the frequency domain, our scheme could realize the effective use of spectrum resources.

\section{Acknowledge}
This work was supported by the National Natural
Science Foundation of China (NSFC) under Grant 62071284,  the National Key Research and Development Program of China under Grant 2022YFB2902304, the Innovation Program of Shanghai Municipal Science and Technology Commission
under Grants 20JC1416400, 21ZR1422400, and 20511106603.

\section{Conclusion} \label{sect:conc}
In this paper, we propose a new system to estimate the high-mobility wireless channel by extracting the dynamic parameters influenced by the mobility velocity and channel frequency. The estimation accuracy is proven to be not relative to the moving velocity and could be used to improve the localization accuracy. We also gave the mathematical proof of our designed system, which demonstrates the enhanced accuracy of channel prediction involving different velocities. Based on channel prediction, localization performance is also improved through the utilization of cross-band CSI information in scenarios. 


\appendices
\section{Proof of Lemma \ref{lem:equ}} 
\label{appen:elbo}
From Eq.(\ref{eq:model}), Eq.(\ref{equ:11}) and Eq.(\ref{eq:begin})-Eq.(\ref{eq:end}), we know that the optimization equation in Problem 2 could be rewritten as:
\begin{eqnarray}
    &&\frac{1}{M}\sum_{m=1}^{M} \|\mathbf{H}_{n'}(m)-\hat{\mathbf{H}}_{n'}(m)\|_2 \\
    &&=\frac{1}{M}\sum_{m=1}^{M} \|\sum_{l = 1}^{L}P_{l}^{n'}(m)-\sum_{l = 1}^{L}\hat{P}_{l}^{n'}(m)\|_2\\
    &&=\frac{1}{M}\frac{1}{L}\sum_{m=1}^{M}\sum_{l=1}^{L} \|{{P}}^{n'}_{l}(m) -\hat{{P}}_{l}^{n'}(m)\|_2\\
    &&=\frac{1}{M}\frac{1}{L}\sum_{m=1}^{M}\sum_{l=1}^{L} \|\Xi -\Xi'\|_2\\
    where,\\
    &&\Xi = {\mathcal{X}}^{-1}_{n'}({\overline{P}^{n'}_{l}}(m)),\\
    &&\Xi' = \hat{\mathcal{X}}^{-1}_{n'}({\tilde{P}^{n'}_{l}}(m))
\end{eqnarray}
In our system, the function ${\mathcal{X}}(\cdot)$ is the function removing dynamic parameters from the channel parameters, and the dynamic parameters are mainly related to the channel frequency and moving velocity.  The Eq.(\ref{eq:DS model}) shows that In our channel model, the function of removing the dynamic channel parameters is to extract the factor from channel information. If we have the condition to estimate the parameters in $\hat{\mathcal{X}}^{-1}_{n'}({\tilde{P}^{n'}_{l}}(m))$ accurately, we could say that $\hat{\mathcal{X}}^{-1}_{n'}(\cdot) \approx {\mathcal{X}}^{-1}_{n'}(\cdot)$, and the equation above could be rewritten as:
\begin{eqnarray}
    &&\frac{1}{M}\frac{1}{L}\sum_{m=1}^{M}\sum_{l=1}^{L} \|\Xi -\Xi'\|_2\\
    &&\approx \frac{\sqrt{N_{n'}^t}}{M}\frac{\sqrt{N_{n'}^t}}{L}\sum_{m=1}^{M}\sum_{l=1}^{L} \|\overline{{P}}^{n'}_{l}(m) -\tilde{{P}}_{l}^{n'}(m)\|_2
\end{eqnarray}
$N_{n'}^t$ is the time-varying factor related to channel frequency, and we extract this factor by FNN. This equation is rational to the optimization equation in Problem 3, which means if the precondition that the parameters were estimated accurately, Problem 1 is equivalent to Problem 3. So we designed a two-stage system and connected them linearly. The first stage is based on Problem 2 to estimate the channel parameters to fulfill the precondition, and the second stage is based on Problem 3 to replace Problem 1 after Problem 2 is solved.

\bibliographystyle{IEEEtran}
\bibliography{IEEEabrv,bb_rf}

\begin{thebibliography}{10}
\providecommand{\url}[1]{#1}
\csname url@samestyle\endcsname
\providecommand{\newblock}{\relax}
\providecommand{\bibinfo}[2]{#2}
\providecommand{\BIBentrySTDinterwordspacing}{\spaceskip=0pt\relax}
\providecommand{\BIBentryALTinterwordstretchfactor}{4}
\providecommand{\BIBentryALTinterwordspacing}{\spaceskip=\fontdimen2\font plus
\BIBentryALTinterwordstretchfactor\fontdimen3\font minus
  \fontdimen4\font\relax}
\providecommand{\BIBforeignlanguage}[2]{{%
\expandafter\ifx\csname l@#1\endcsname\relax
\typeout{** WARNING: IEEEtran.bst: No hyphenation pattern has been}%
\typeout{** loaded for the language `#1'. Using the pattern for}%
\typeout{** the default language instead.}%
\else
\language=\csname l@#1\endcsname
\fi
#2}}
\providecommand{\BIBdecl}{\relax}
\BIBdecl

\bibitem{6GISAC}
F.~Liu, Y.~Cui, C.~Masouros, J.~Xu, T.~X. Han, Y.~C. Eldar, and S.~Buzzi,
  ``Integrated sensing and communications: Toward dual-functional wireless
  networks for 6g and beyond,'' \emph{IEEE Journal on Selected Areas in
  Communications}, vol.~40, no.~6, pp. 1728--1767, 2022.

\bibitem{10157900}
K.~NonAlinsavath, V.~Khieovongphachanh, P.~Southisombath, A.~Chaisang,
  S.~Phomkeona, K.~Luangxaysana, S.~Suwan, and S.~Promwong, ``Location context
  awareness system for specific positioning based on received signal strength
  for android platform system,'' in \emph{2023 9th International Conference on
  Engineering, Applied Sciences, and Technology (ICEAST)}, 2023, pp. 93--96.

\bibitem{8275781}
S.~Xu and W.~Chou, ``An improved indoor localization method for mobile robot
  based on wifi fingerprint and amcl,'' in \emph{2017 10th International
  Symposium on Computational Intelligence and Design (ISCID)}, vol.~1, 2017,
  pp. 324--329.

\bibitem{7990593}
X.~Guo and N.~Ansari, ``Localization by fusing a group of fingerprints via
  multiple antennas in indoor environment,'' \emph{IEEE Transactions on
  Vehicular Technology}, vol.~66, no.~11, pp. 9904--9915, 2017.

\bibitem{8423070}
Y.~Xie, Z.~Li, and M.~Li, ``Precise power delay profiling with commodity
  wi-fi,'' \emph{IEEE Transactions on Mobile Computing}, vol.~18, no.~6, pp.
  1342--1355, 2019.

\bibitem{souzandeh2023classification}
N.~Souzandeh and J.~Pourahmadazar, ``Classification of frequency bands in the
  joint millimeter-wave sensing and communication system (jscs): A
  comprehensive analysis,'' Ph.D. dissertation, Institut National de la
  Recherche Scientifique [Qu{\'e}bec], 2023.

\bibitem{yuan2023variational}
R.~Yuan, K.~Huang, P.~Yang, and S.~Zhang, ``A variational auto-encoder enabled
  multi-band channel prediction scheme for indoor localization,'' \emph{arXiv
  preprint arXiv:2309.12200}, 2023.

\bibitem{8424581}
X.~Shen, Y.~Liao, X.~Dai, M.~Zhao, K.~Liu, and D.~Wang, ``Joint channel
  estimation and decoding design for 5g-enabled v2v channel,'' \emph{China
  Communications}, vol.~15, no.~7, pp. 39--46, 2018.

\bibitem{shi2021unified}
Q.~Shi, Y.~Liu, S.~Zhang, S.~Xu, and V.~K. Lau, ``A unified channel estimation
  framework for stationary and non-stationary fading environments,'' \emph{IEEE
  Transactions on Communications}, vol.~69, no.~7, pp. 4937--4952, 2021.

\bibitem{10128162}
S.~Cai, L.~Chen, Y.~Chen, H.~Yin, and W.~Wang, ``Pulse-based isac: Data
  recovery and ranging estimation for multi-path fading channels,'' \emph{IEEE
  Transactions on Communications}, vol.~71, no.~8, pp. 4819--4838, 2023.

\bibitem{10000625}
Y.~Wan, A.~Liu, Q.~Hu, M.~Zhang, and Y.~Cai, ``A two-stage global estimation
  scheme for multiband delay estimation in wireless localization,'' in
  \emph{2022 IEEE Global Communications Conference (GLOBECOM)}, 2022, pp.
  735--740.

\bibitem{article1}
Z.~Zhou, C.-X. Wang, L.~Zhang, J.~Huang, L.~Xin, E.-H. Aggoune, and Y.~Miao,
  ``A novel sage algorithm for estimating parameters of wideband spatial
  non-stationary wireless channels with antenna polarization,'' \emph{IEEE
  Transactions on Antennas and Propagation}, vol.~PP, pp. 1--1, 09 2023.

\bibitem{10.1145/3210240.3210314}
\BIBentryALTinterwordspacing
K.~Qian, C.~Wu, Y.~Zhang, G.~Zhang, Z.~Yang, and Y.~Liu, ``Widar2.0: Passive
  human tracking with a single wi-fi link,'' in \emph{Proceedings of the 16th
  Annual International Conference on Mobile Systems, Applications, and
  Services}, ser. (MobiSys).\hskip 1em plus 0.5em minus 0.4em\relax New York,
  NY, USA: Association for Computing Machinery, 2018, p. 350–361. [Online].
  Available: \url{https://doi.org/10.1145/3210240.3210314}
\BIBentrySTDinterwordspacing

\bibitem{dempster1977maximum}
A.~P. Dempster, N.~M. Laird, and D.~B. Rubin, ``Maximum likelihood from
  incomplete data via the em algorithm,'' \emph{Journal of the royal
  statistical society: series B (methodological)}, vol.~39, no.~1, pp. 1--22,
  1977.

\bibitem{Yin2016PerformanceCO}
\BIBentryALTinterwordspacing
X.~F. Yin, L.~Ouyang, and H.~Wang, ``Performance comparison of sage and music
  for channel estimation in direction-scan measurements,'' \emph{IEEE Access},
  vol.~4, pp. 1163--1174, 2016. [Online]. Available:
  \url{https://api.semanticscholar.org/CorpusID:27030716}
\BIBentrySTDinterwordspacing

\bibitem{6901837}
X.~Yin, Y.~He, Z.~Song, M.-D. Kim, and H.~K. Chung, ``A
  sliding-correlator-based sage algorithm for mm-wave wideband channel
  parameter estimation,'' in \emph{The 8th European Conference on Antennas and
  Propagation (EuCAP 2014)}, 2014, pp. 625--629.

\bibitem{fleury1999channel}
B.~H. Fleury, M.~Tschudin, R.~Heddergott, D.~Dahlhaus, and K.~I. Pedersen,
  ``Channel parameter estimation in mobile radio environments using the sage
  algorithm,'' \emph{IEEE Journal on selected areas in communications},
  vol.~17, no.~3, pp. 434--450, 1999.

\bibitem{934374}
Eberhart and Y.~Shi, ``Particle swarm optimization: developments, applications
  and resources,'' in \emph{Proceedings of the 2001 Congress on Evolutionary
  Computation (IEEE Cat. No.01TH8546)}, vol.~1, 2001, pp. 81--86 vol. 1.

\bibitem{9740477}
L.~Liu, J.~Jin, and S.~Xiong, ``Performance optimization of polarized mimo
  relay channels based on the cost 2100 channel model,'' \emph{IEEE Antennas
  and Wireless Propagation Letters}, vol.~21, no.~6, pp. 1188--1192, 2022.

\bibitem{hansen2016cma}
N.~Hansen, ``The cma evolution strategy: A tutorial,'' \emph{arXiv preprint
  arXiv:1604.00772}, 2016.

\bibitem{alkhateeb2019deepmimo}
A.~Alkhateeb, ``Deepmimo: A generic deep learning dataset for millimeter wave
  and massive mimo applications,'' \emph{arXiv preprint arXiv:1902.06435},
  2019.

\end{thebibliography}

\end{document}